# Robust Surface Reconstruction from Orthogonal Slices

**Radek Sviták[1], Václav Skala[2]**

Department of Computer Science and Engineering,
University of West Bohemia in Pilsen, Univerzitní 8, 306 14 Plzeň, Czech Republic
E-mail: rsvitak@kiv.zcu.cz

## Abstract

The surface reconstruction problem from sets of planar parallel slices representing cross sections through 3D objects is presented. The final result of surface reconstruction is always based on the correct estimation of the structure of the original object. This paper is a case study of the problem of the structure determination. We present a new approach, which is based on considering mutually orthogonal sets of slices. A new method for surface reconstruction from orthogonal slices is described and the benefit of orthogonal slices is discussed too. The properties and sample results are presented as well.

## 1. Introduction

The crucial task of the surface reconstruction from slices is a correct estimation of the original object structure, i.e. the solution of the contour correspondence problem. Most of the existing methods simply consider the overlap of contours in a pair of consecutive parallel slices as the only correspondence criterion. Therefore, they produce unacceptable structure estimation when the angle between the axis of the object and the normal of the slices increases.

Higher density of slices can help to solve this problem, but it is not always possible because of the resolution limit of the scanning device, etc. It is obvious that other slices in non-parallel planes offer an additional information. In this paper we will concentrate on the benefit of orthogonal slices for the reconstruction process. In comparison to the existing methods, our currently achieved results show, that for a set of objects the resultant surface is significantly more accurate with respect to the similarity to the original surface.

The concept of the new proposed method is presented and results of comparisons with the existing methods are discussed as well.

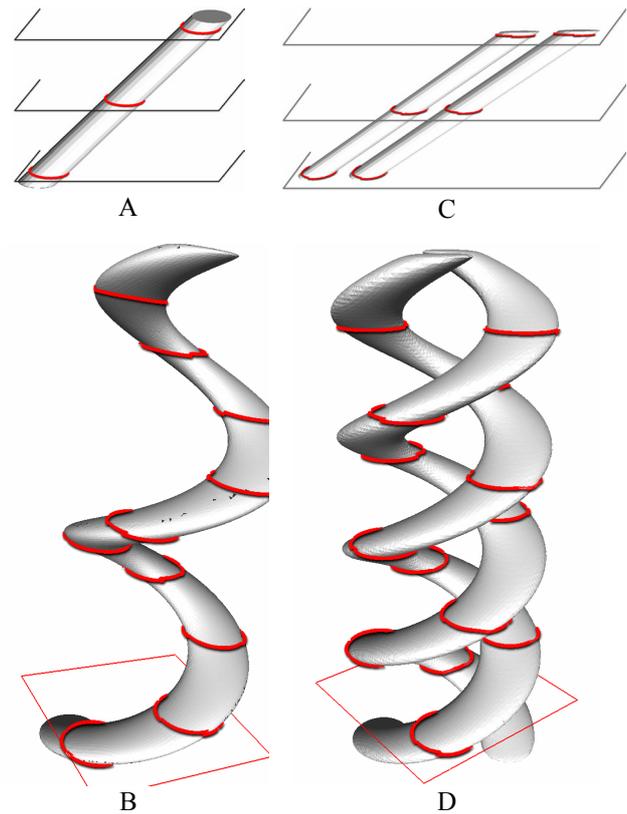

Figure 1: Problematic cases when solving the contour correspondence problem. Expected problems using the overlapping criterion: A, B, C, D; generalized cylinders: B, D; MST: C, D; Reeb graph based methods: D.

This work is was supported by the Ministry of Education of the Czech Republic – projects:
[1]FRVŠ 1348/2004/G1
[2]MSM 235200005

## 2. Brief survey of existing methods

Several methods for surface reconstruction from slices have been developed since about 1970. In this section we will classify them according to their approach to solving the contour correspondence problem. For more extensive study of the existing methods from the other viewpoints, see [2, 4, 6, 7].

The simplest methods estimate the contour correspondence locally between each consecutive pair of contours. Typically, contours that overlap each other are considered as correspondent. This works if the density of slice is high, i.e. the distance between slices is low, and the axis of the input object is nearly perpendicular to the slices planes.

A more advanced method uses *generalized elliptical cylinder* to solve the correspondence problem [1, 11, 12]. Contours are first classified as elliptical or complex by determining how well the vertices of their perimeter can be fit by an ellipse. If the fit is too poor, a contour is classified as *complex*, and can not be incorporated into an elliptical cylinder. Then the ellipses are grouped to the cylinders. When as many contours as possible have been organized into cylinders, then the algorithm uses the geometric relationship between cylinders to group them into objects. This method is most useful for elongated smooth objects with roughly elliptical cross section.

Apparently the best existing approaches that have been published are two graph-based methods. The first of them presented by Skinner [10] computes a *minimum spanning tree* based on contour shape and position. In the first step a graph is constructed by representing each contour as a node and connecting each node to all nodes representing contours in adjacent sections. The best fitting ellipse is computed for each contour. The cost of an edge of the graph relies on the mutual position and size of two ellipses:

$$c(i,j) = (x_i - x_j)^2 + (y_i - y_j)^2 + (a_i - a_j)^2 + (b_i - b_j)^2,$$

where $(x_i, y_i, z_i)$, $(x_j, y_j, z_j)$ represent the centers of the ellipses of contours $i$, $j$, respectively, and $a_i$, $b_i$, $a_j$, $b_j$ are their major and minor axis lengths.

The minimum spanning tree computed for the graph represents the solution to the correspondence problem. The method works well for naturally tree-structured objects, the main limitation is its inability to solve the correspondence problem correctly for general graph topologies, e.g. genus > 0.

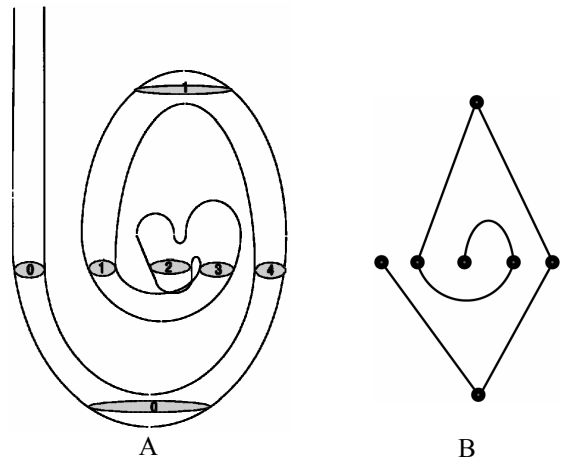

Figure 2: A) Data set of slices of the cochlea. Using the Reeb graph it is possible to detect and represent the right contour correspondence. The advantage consists in the possibility of considering the correspondence among contours of one slice (B). Taken from [8].

The second graph based method presented by Shinagawa [8, 9] uses surface coding based on Morse Theory to construct a *Reeb graph* [14] representing the contour connectivity. Each contour represents a node in the graph, edges of the graph represent the contour correspondence relation. Edges are added to the graph in the manner to avoid making connections that would result in a surface that is not a 2-manifold. For each pair of contours that can be legally connected, a weight function is evaluated, and its value is used to establish a priority for connecting that pair of contours. The algorithm proceeds by making the highest priority connections in regions where the number of contours in each section does not change, and then adds connections in order of decreasing priority with respect to the *a priori knowledge* of the *number of connected components* and the *topological genus*.

It is necessary to note that all the existing solutions just *estimate* the contours correspondence, i.e. the structure of the original object, should be emphasized. In Figure 1 there are some typical example data sets to illustrate capabilities of the approaches mentioned in this section.

## 3. Orthogonal slices

One set of parallel planar slices is one of the well-known boundary representations of a 3D object. Usually the planes of such slices set are perpendicular to the z-axis, and thus called *z-slices*.

If we slice an object by more then one set of parallel slices and moreover when these sets are mutually orthogonal, we get orthogonal sets of slices. Consider now that we have *z-slices*, *x-slices* and *y-slices* of an object, see Figure 3. Note, that the contours are supposed to be polygonal, oriented the way that when looking from the positive direction of the given slices set axis, the contours have the interior on its left side and the exterior on the right side, see Figure 4.

**3.1 Contour correspondence**

The main advantage of orthogonal slices consists in the approach how the contour correspondence can be determined. It is important to emphasize that two orthogonal contours which intersect each other comes aparently from one and the same surface component of the input object. It means that the *intersection of contours* is very important since it provides accurate information about the correct structure, see Figure 5.

It is obvious that if the slices in the orthogonal sets sample the object *sufficiently*, then the intersections of contours from the orthogonal slices identify the correspondence relation accurately, i.e. the correct structure of the original object.

## 4. The algorithm

The planes of slices divide space into a set of *spatial cells* of a *spatial grid M*. In Figure 3 can be seen three mutually orthogonal planes of grid *M*. We distinguish two kinds of cells of *M*, the *surface-crossing* and the *surface-passing* cells. There are parts of contours on some sides of a surface-crossing cell, which means that the resultant surface intersects the cell, see Figure 6.

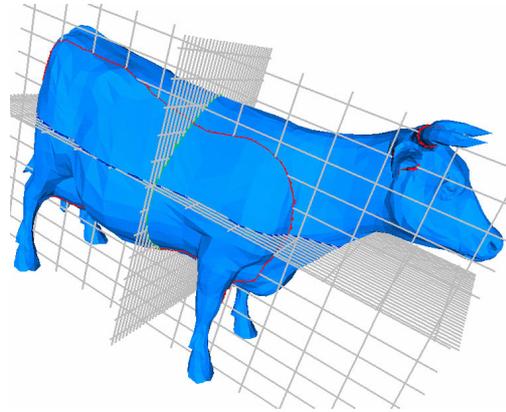

Figure 3: An example of three orthogonal slices sets.

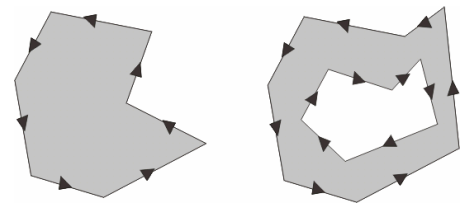

Figure 4: Correct contour orientation.

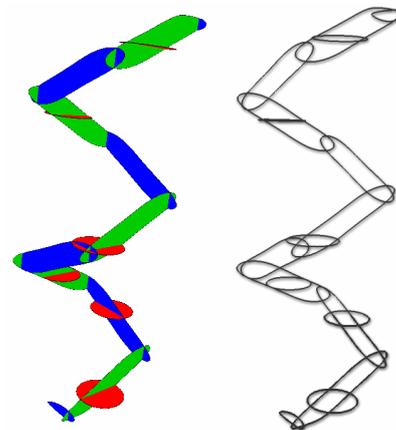

Figure 5: The mutual crossings of orthogonal contours define the correspondence relation.

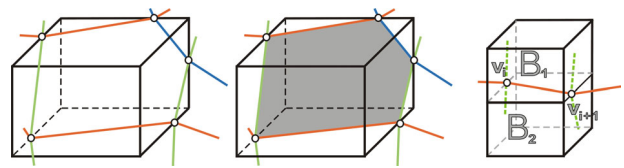

Figure 6: A surface-crossing cell. Parts of contours on the sides of the cell together with node points form spatial polygons. Node points are denoted as white circles. Each edge of G is adjacent with two cells of *M*.

The intersection of two orthogonal slices consisting of curvilinear contours is a set of points and we call them *node points*, see Figure 7. Now we focus on surface-crossing cell. An important observation is that parts of input contours and the node points form *spatial polygons*. Each such polygon is enclosed in a surface-crossing cell, its patch is part of the resultant surface, see Figure 6.

### 4.1 The correspondence problem

At this moment we suppose that the correspondence of contours is identified sufficiently by the intersections of orthogonal contours as it has been discussed in section 3.1.

Consider the intersection of two contours as the relation of correspondence. Note that the number of components of a graph constructed of such a relation corresponds to the number of disjoint components of the resultant surface.

### 4.2 Node points computation

A node point is geometrically the intersection of two contours. Topologically it is the representation of a contour correspondence relation. It holds that each node point must lie on the edge of the grid $M$. Since the contours are supposed as polygonal curves, we cannot compute the intersections of two orthogonal sections directly. We obtain them in two phases.

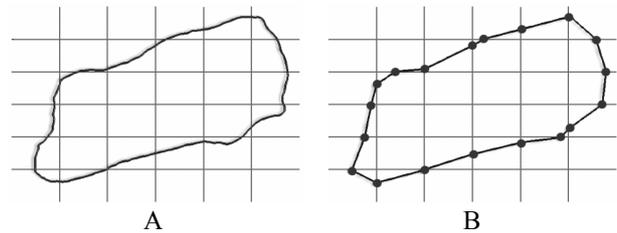

Figure 7: A) An input contour, the lattice represents positions of orthogonal slices planes. B) The contour formed by its node points (black spots).

In the first step intersections of each contour and the grid $M$ are computed. These intersections are added among the current contour vertices on the appropriate position. They are *registered on the corresponding edge* of the grid $M$ simultaneously. Our algorithm works on the same principle as the Cohen-Sutherland's line clipping algorithm [3].

An intersection of a slice plane and all other orthogonal slice planes forms a lattice with cells, see Figure 7. Each node point arises as the intersection of a contour and a side of a cell. Since the contour is supposed to be polygonal, a node point is simply computed as an intersection of two segments. Singular cases when a contour crosses a cell at its corner are handled separately [13].

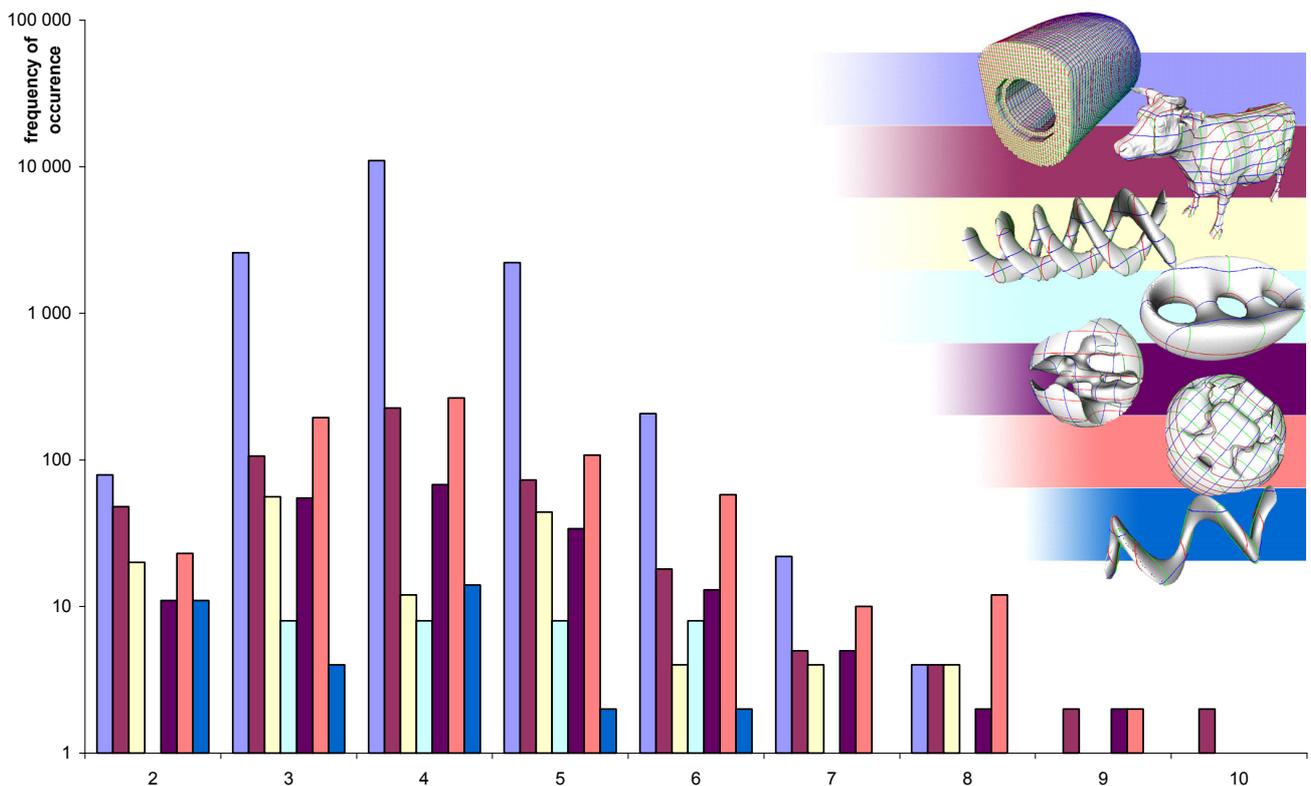

Figure 8: Polygon size (number of edges) histogram.

In the second step the node point construction is completed. The *correspondent vertex*, which is a member of the orthogonal contour and also a member of the same edge of *M*, must be found. As it was said before it is done very fast searching the auxiliary registrations of contour intersections on the appropriate edge of *M*. Each two nearest intersections coming from orthogonal contours registered on an edge of *M* are qualified as correspondent vertices building together a node point.

### 4.3. Constructing the surface

Now suppose graph *G*, whose set of vertices consists of a set of the node points and whose edges represent the parts of contours between two node vertices. Note that the geometrical shape of the edges still corresponds to the appropriate parts of contours. Now the task is to find such cycles of graph *G*, which have the property that their geometrical representation lies within one cell of *M*. Those cycles represents spatial polygons that lie on the surface.

We suppose each edge *e* of *G* is adjacent with cells $B_1$ and $B_2$, see Figure 6. Each cell from $\{B_1, B_2\}$ includes one cycle *c* of our interest, which is adjacent with *e* (that results from the consideration of 2-manifold objects). The circle *c* represents the spatial polygon being searched. Thus for each *e* two cycles $c^e_{B_1}$, $c^e_{B_2}$ must be searched and then polygons $p_{c_1}$, $p_{c_2}$ correspondent to those cycles are constructed.

As soon as all polygons are obtained, we can start to patch them. We can use any arbitrary patching technique. Note that the number of sides of such polygon can be high, but in cases of our data sets it is in range 2 – 10, see the graph in Figure 8.

The proposed method starts with finding a suitable point in the center of each polygon. Then using the center point each polygon is divided into set of quadrilaterals, which are easier to patch, see Figure 9.

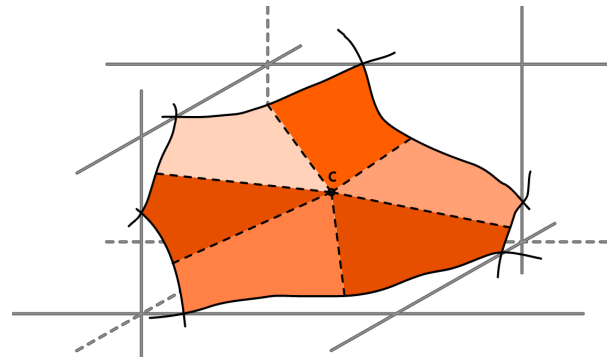

Figure 9: Partition of a generic polygon in the set of quadrilaterals. Requires the central point *C* determination.

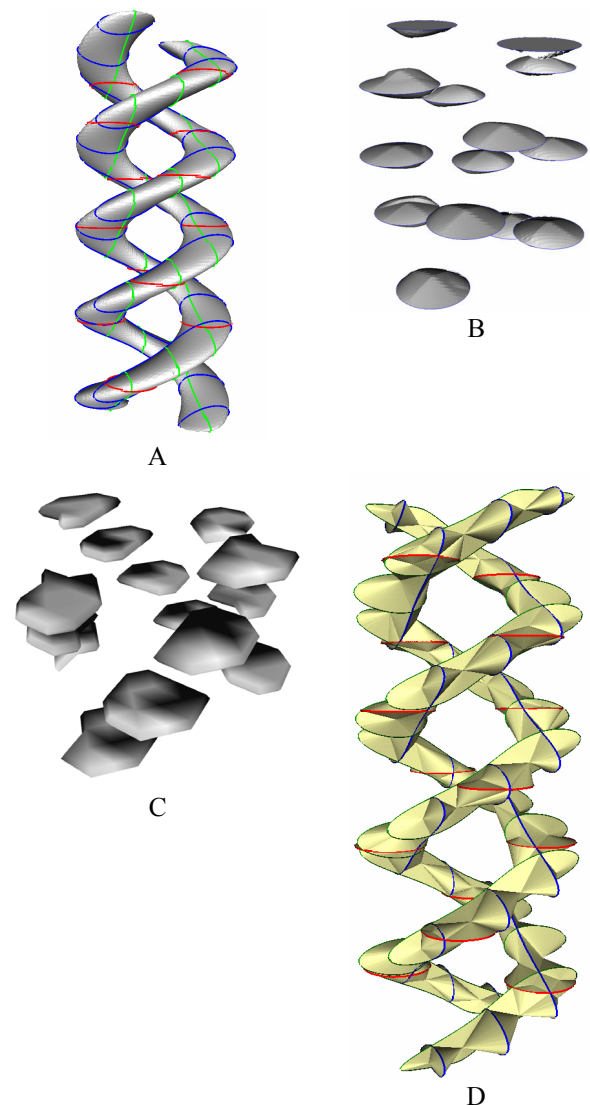

Figure 10: Results of the surface reconstruction. A) An input data set (courtesy of Martin Čermák), B) VTK surface reconstruction from slices class, C) A common volume based method, D) Proposed method for surface reconstruction from orthogonal slices.

## 5. Results

All the problematic data sets mentioned in section 1 and many more have been processed using:
- surface reconstructing from slices class from VTK,
- a common volume based method; see [5] for more details,
- our proposed method for surface reconstruction from orthogonal slices.

The results of the reconstruction of one data set are illustrated in Figure 10, the complete documentation and experimental results can be found at http://herakles.zcu.cz/research/slices.

## 6. Conclusion and further research

Our current research proves that the advantages of orthogonal slices in the process of surface reconstruction are significant. There is a set of objects for which the orthogonal slices are almost the only way to reconstruct them correctly.

The proposed method supposes that the object is sampled well enough, so that the number of components of the correspondence graph $G$ equals to the number of disjoint components of the original surface.

The main point of our further research is the solution of problems caused by under-sampling, i.e. to deal with data sets that do not sample the input object sufficiently. Furthermore we would like to study the influence of contour inaccuracy on the node point computation.

## References

[1] Bresler, Y., Fessler, J.A., Macovski, A.: A Bayesian approach to reconstruction from incomplete projections of a multiple object 3D domain. IEEE Trans. Pat. Anal. Mach. Intell., 11(8):840-858, August 1989.

[2] Cong, G., Parvin, B.: Robust and efficient surface reconstruction from contours. The Visual Computer, (17):199-208, 2001

[3] Foley, J. D., van Dam, A., Feiner, S. K. and Hughes, J. F., Computer Graphics: Principles and Practice, Addison-Wesley, 1990.

[4] Jones, M., Chen, M.: A new approach to the construction of surfaces from contour data. Computer Graphics Forum (13): 75-84, 1994

[5] Klein, R., Schilling, A.: Fast Distance Interpolation for Reconstruction of Surfaces from Contours. In proceedings of Eurographics '99, Short Papers and Demos, September 1999.

[6] Meyers, D.: Multiresolution tiling. In Proceedings, Graphics Interface '94, pages 25-32, Banff, Alberta, May 1994.

[7] Meyers, D.: Reconstruction of Surfaces From Planar Contours. PhD thesis, University of Washington, 1994.

[8] Shinagawa, Y., Kunii, T.L.: Constructing a Reeb graph automatically from cross sections. IEEE Comuter Graphics and Applications, 11(6): 44-51, November 1991.

[9] Shinagawa, Y., Kunii, T.L., Kergosien, Y.L.: Surface coding based on Morse theory. IEEE Comuter Graphics and Applications, 11(5): 66-78, September 1991.

[10] Skinner, S.M.: The correspondence problem: Reconstruction of objects from contours in parallel sections. Master's thesis, Department of Computer Science and Engineering, University of Washington, 1991.

[11] Soroka, B.I.: Understanding Objects From Slices: Extracting Generalised Cylinder Descriptions From Serial Sections. PhD thesis, University of Kansas Dept of Computer Science, March 1979. TR-79-1.

[12] Soroka, B.I.: Generalized cones from serial sections. Computer Graphics and Image Processing, (15): 54-166, 1981.

[13] Svitak, R., Skala, V.: Surface Reconstruction from Orthogonal Slices, ICCVG 2002, Zakopane, Poland, 2002

[14] Wood, Z. J.: Computational Topology Algorithms For Discrete 2-Manifolds. California Institute of Techology, PhD Thesis, May 2003